\documentclass[12pt]{iopart}
\usepackage{graphicx}

\begin{document}

%\title[Strange hadrons at intermediate $p_{T}$]{Strange hadrons at
%  intermediate transverse momentum: evidence for a deconfined
%  quark-gluon phase in Au+Au collisions at $\sqrt{s_{_{NN}}}=200$~GeV
%}

\title[Quark and Gluon Degrees of Freedom]{Evidence from Identified
  Particles for Active Quark and Gluon Degrees of Freedom }

\author{Paul Sorensen}

\address{Brookhaven National Laboratory, P.O. Box 5000, Upton, NY 11973}
\ead{prsorensen@bnl.gov}

\begin{abstract} 
Measurements of intermediate $p_T$ ($1.5<p_T<5.0$~GeV/c) identified
particle distributions in heavy ion collisions at SPS and RHIC
energies display striking dependencies on the number of constituent
quarks in the corresponding hadron. One finds that elliptic flow at
intermediate $p_T$ follows a constituent quark scaling law as
predicted by models of hadron formation through coalescence. In
addition, baryon production is also found to increase with event
multiplicity much faster than meson production. The rate of increase
is similar for all baryons, and seemingly independent of mass. This
indicates that the number of constituent quarks determines the
multiplicity dependence of identified hadron production at
intermediate $p_T$. We review these measurements and interpret the
experimental findings.
\end{abstract}

%\keyword{Nuclear Modification, Elliptic Flow} 


\submitto{\JPG} 

\section{Introduction}\label{intro}
Physicists at the Relativistic Heavy Ion Collider (RHIC) have made
several unexpected and potentially interesting
observations~\cite{Adcox:2004mh}. Measurements relating to baryon
production in the intermediate transverse momentum region ($1.5<p_T<5$
GeV/c) are one of those
\cite{Adler:2003kg,Adams:2003am,Sorensen:2004wg}. In p+p collisions at
$p_T=3$~GeV/c, one baryon is produced for every three mesons (1:3). In
central Au+Au collisions however, baryons and mesons are created in
nearly equal proportion (1:1). At this same $p_T$, the elliptic
anisotropy ($v_2$) of baryons is also 50\% larger than meson elliptic
flow: demonstrating that baryon production is also enhanced in the
direction of the impact vector between the colliding nuclei
(in-plane)~\cite{Adams:2003am,Adams:2004bi}. The enhancement persists
up to $p_T=5.5$~GeV/c.

This effect is generic to many systems; baryon production is enhanced
in systems that yield larger multiplicity and have the potential for
an increase in multi-parton effects. Several possible explanations for
the baryon enhancement are commonly considered:
%\begin{description} 
%\item[Hadronization ] multi-quark or gluon hadron formation processes such as
%  coalescence or recombination.
%\item[Gluon Junctions ] topological gluon configurations carrying baryon number.
%\item[Flow ] collective motion amongst more massive baryons that
%  populates the higher $p_T$ regions of baryon $p_T$ spectra.
%\end{description}
\begin{quote} 
Multi-quark or gluon processes during hadron
formation---\emph{coalescence}~\cite{reco}.

Gluon configurations that carry baryon number---\emph{baryon
  junctions}~\cite{junctions,vitevjunctions}.

Collective motion amongst more massive baryons that populates the
higher $p_T$ regions of baryon $p_T$
spectra---\emph{flow}~\cite{hydro}.
\end{quote}
Coalescence models have garnered particular attention because they
seem to provide a natural explanation for the constituent-quark-number
scaling that has been observed in $v_2$ measurements. They also relate
hadronic observables to a pre-hadronic stage of interacting quarks and
gluons. As such, they touch on questions central to the heavy-ion
physics program: deconfinement and chiral symmetry restoration.  Given
the potential physics benefits that can be derived from these
measurements, experimentalists at RHIC and SPS have endeavored to
extend their abilities to identify hadron types up to higher $p_T$
regions~\cite{st_pid}.  Here I review those measurements which provide
evidence for active quark and gluon degrees-of-freedom and discuss
implications of the baryon enhancement on pion and non-photonic
electron spectra.

\section{Baryon to Meson Ratios}

\begin{figure}[htb]
\centering\mbox{
\vspace{-15pt}
\includegraphics[width=1.0\textwidth]{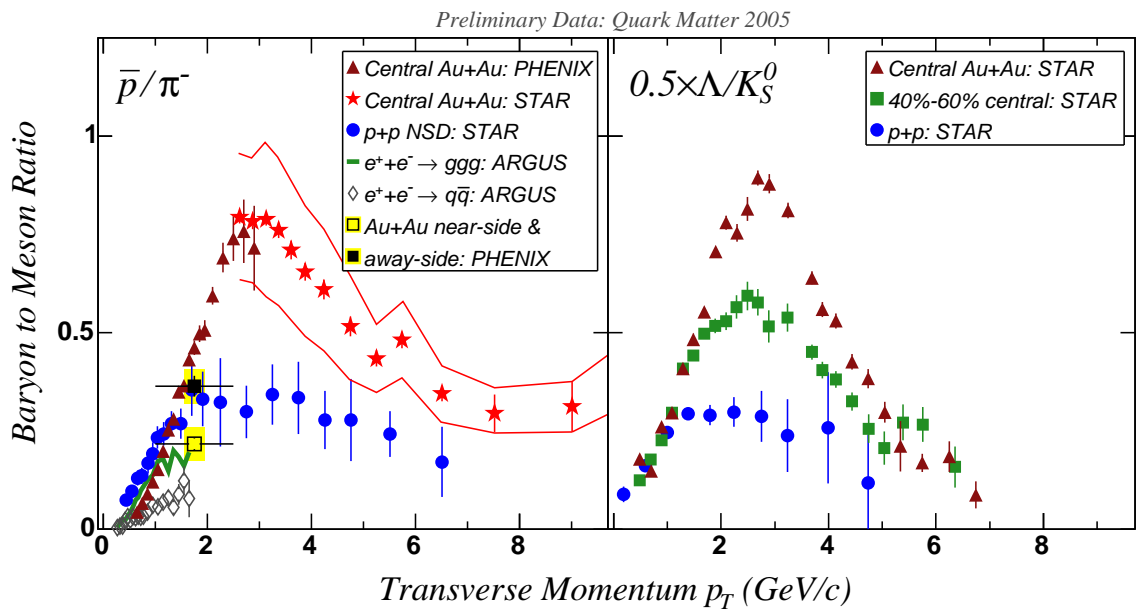}}
\vspace{-15pt}
\caption[]{ Left panel: the $\overline{p}/\pi^-$ ratio at middle
  rapidity for central Au+Au and p+p collisions at
  $\sqrt{s_{_{NN}}}=200$~GeV. ARGUS measurements of the proton to pion
  ratio in $e^++e^-$ collisions at $\sqrt{s}=10$~GeV are shown for two
  classes of events: $e^+e^- \rightarrow \Upsilon \rightarrow ggg$ and
  continuum events dominated by $e^+e^- \rightarrow
  q\overline{q}$. Measurements of the proton to pion ratio made for
  particles associated with a trigger hadron ($p_T>2.5$) are also
  shown~\cite{Sickles:2004jz}. Right panel: $\Lambda/K_S^0$ in central
  Au+Au, mid-peripheral Au+Au and minimum-bias p+p
  collisions~\cite{LoverK}. Values are scaled by 0.5.}
\label{ratios}
\end{figure}

Figure~\ref{ratios} (left panel) shows the $\overline{p}/\pi^-$ ratio
measured in $e^++e^-$~\cite{argus}, p+p~\cite{olga}, and
Au+Au~\cite{Adler:2003kg,olga} collisions. The right panel shows the
$\Lambda/K^0_S$ ratio for p+p, mid-peripheral Au+Au, and central
Au+Au collisions scaled by 0.5~\cite{LoverK}. In central Au+Au collisions,
$\overline{p}/\pi^-$ reaches a maximum value of nearly 1 at
$p_T\approx3$~GeV/c. The baryon junction calculations in
Ref.~\cite{vitevjunctions} predict that the $p_T$ value where the
$B/M$ ratio is at its maximum will increase with collision
centrality. This prediction can be compared to the $\Lambda/K_S^0$
data. Measurements are still not precise enough, however, to confirm
nor disprove this prediction. Figure~\ref{ratios} demonstrates that
the baryon enhancement in Au+Au collisions is part of a
systematic trend.  Baryon production is also enhanced in
$\sqrt{s}=10$~GeV $e^++e^-$ collisions when $e^+e^- \rightarrow
\Upsilon \rightarrow ggg$ events are compared to continuum $e^+e^-
\rightarrow q\overline{q}$ events~\cite{argus}. From those
measurements the question arose: \textit{is the enhancement related
  to multi-parton topological effects or a difference between quark
  and gluon fragmentation?} Since $e^+e^- \rightarrow \Upsilon
\rightarrow ggg$ is a purely gluonic process, the observation that the
$\overline{p}/\pi^-$ ratio is even larger in Au+Au collisions
indicates that multi-parton topological effects drive the enhancement
(baryon junctions or coalescence).

\subsection{Azimuthal Dependence: $v_2$}

\begin{figure}[htb]
\centering\mbox{
\vspace{-10pt}
\includegraphics[width=0.8\textwidth]{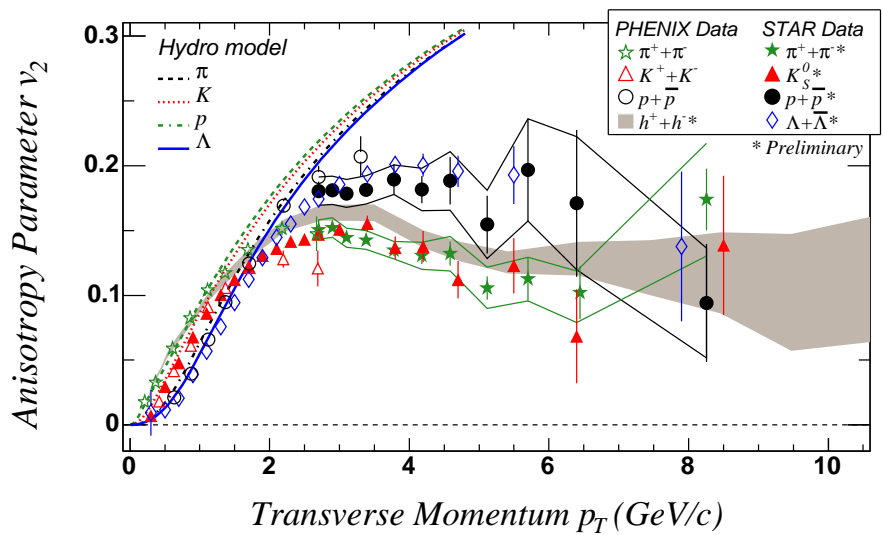}}
\vspace{-0pt}
\caption[]{ Elliptic flow measurements at middle rapidity from
  minimum-bias Au+Au collisions at $\sqrt{s_{NN}}=200$~GeV. The bands
  around the STAR preliminary measurements of pions and protons
  represent systematic uncertainties mostly from non-flow
  correlations. The PHENIX measurements are made by correlating
  hadrons at middle rapidity with an event-plane measured using hadrons
  at $3.1<|\eta|<4.0$. }
\label{v2}
\end{figure}

$v_2$ is a sensitive measure to distinguish between possible
interpretations of the hadron type dependence of particle spectra.
Figure~\ref{v2} shows preliminary measurements of $v_2$ with a minimum
bias centrality selection in Au+Au collisions at
$\sqrt{s_{_{NN}}}=200$~GeV~\cite{Oldenburg:2005er,Masui:2005aa}. The
curves show $v_2$ for pions, kaons, protons and $\Lambda$s from a
hydrodynamic calculation~\cite{hydro}. At $p_T<1.0$~GeV/c, the mass
ordering suggests that $v_2$ in that region results from collective
motion. The hydrodynamic calculations capture some of the general
features of the data in this region. At much higher $p_T$ it is
expected that $v_2$ will be developed via energy loss by fast partons
as they traverse the medium created in the collisions. Calculations
suggest that $v_2$ from this mechanism should be less than
10\%~\cite{dedx}. It is also expected that all hadrons will have
similar $v_2$ values where partonic energy loss is the dominant source
of $v_2$. The magnitude and particle-type dependence of $v_2$ seen for
$p_T<6$~GeV/c suggests that other mechanisms contribute to the
development of $v_2$ up to 6--7 GeV/c. This observation is consistent
with results from a parton cascade model predicting that the effects
of flow may persist up to $p_T=7$~GeV/c~\cite{Molnar:flow}. At
$p_T=7$~GeV/c, within errors, the particle-type dependence of
$v_2$ disappears and the $v_2$ measurements are consistent with
expectations from energy loss models~\cite{dedx}.

\subsection{Quark Number Dependence}

%\begin{figure}[htb]
%\centering\mbox{
%\includegraphics[width=0.65\textwidth]{figure_2.eps}}
%\caption[]{ Quark number scaled elliptic flow for preliminary PHENIX
%  and STAR identified hadrons (top panel). To test the scaling
%  hypothesis, a polynomial is fit to all data points. $(B-M)/(B+M)$
%  represents the deviation of quark number scaled baryon $v_2$ and
%  quark number scaled meson $v_2$. Model predictions for deviations
%  from quark number scaling are shown as curves in the bottom panel. }
%\label{scaling}
%\end{figure}
\begin{figure}[htb]
\parbox{0.55\textwidth}{
\centering
\includegraphics[width=0.72\textwidth]{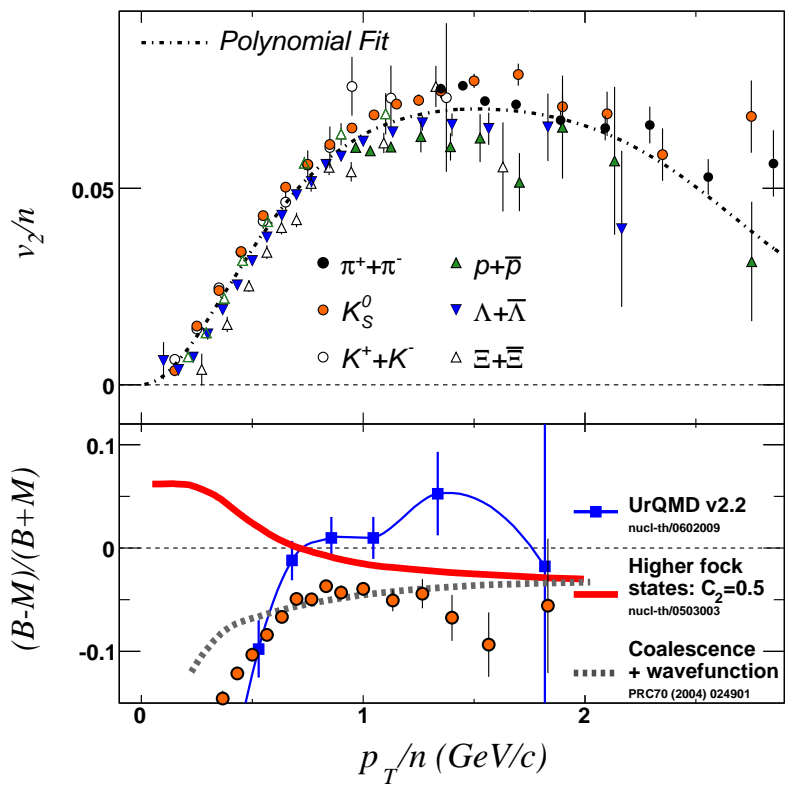}
\vspace{-30pt}
}
\parbox{0.45\textwidth}{
  \caption[]{ Top panel: Quark number scaled elliptic flow for
    identified hadrons (Preliminary)~\cite{Oldenburg:2005er,Masui:2005aa}. A
      polynomial is fit to all data points. Bottom panel. The
      difference between quark number scaled baryon $v_2$ and quark
      number scaled meson $v_2$ divided by the sum:
      $(B-M)/(B+M)$. Here the ratio is taken using
      $\Lambda+\overline{\Lambda}$ and $K_S^0$. Model predictions are
      also shown on the lower panel. The model that provides the best
      description of the data is the recombination model that takes
      the wave-function of the hadron into account. Kaon and $\Lambda$
      data were used for the UrQMD model comparison. }
\label{nqscaling}}
\end{figure}

Hadronization by coalescence or recombination of constituent quarks is
thought to explain many features of hadron production in the
intermediate $p_T$ region ($1.5<p_T<5$~GeV/c)~\cite{reco}. These
models find that at intermediate $p_T$, $v_2$ may follow a
quark-number ($n_q$) scaling with $v_2(p_T/n_q)/n_q$ for all hadrons
falling on one curve. This scaling behavior has been observed in
Au+Au collisions at 200 GeV~\cite{Adams:2004bi}. More sophisticated
theoretical considerations have led to predictions of fine structure
in quark-number scaling---with predictions for a baryon $v_2/n_q$
being smaller than meson $v_2/n_q$~\cite{Greco:2004ex,fock}.
Figure~\ref{nqscaling} (top panel) shows $v_2$ vs $p_T$ for identified
particles, where $v_2$ and $p_T$ have been scaled by $n_q$. A
polynomial function is fit to the scaled values. The deviations from
the scaling are shown in the bottom panel by plotting the difference
between the scaled baryon $v_2$ and the scaled meson $v_2$ divided by
the sum $(B-M)/(B+M)$. Theoretical predictions for this value are also
shown. The model that compares best to data is a coalescence model
that includes the effect of quark momentum distributions inside the
hadron (\textit{Coalescence +
  wavefunction})~\cite{Greco:2004ex}. Accounting for the substructure
of constituent quarks (\textit{higher fock states})~\cite{fock} leads
to a negative $(B-M)/(B+M)$ ratio but with a smaller magnitude than
observed in the data. The hadron/string model UrQMD yields $v_2$
values smaller by a factor of two than experimental data. The model
does, however, find a scaling with quark number. In this model the
scaling originates from the additive quark model for hadronic
cross-sections. It should be noted, when considering comparisons to
any models, that the systematic uncertainties on these preliminary
measurements have not yet been specified.

\begin{figure}[htb]
\vspace{-10pt}
\centering\mbox{
\includegraphics[width=1.0\textwidth]{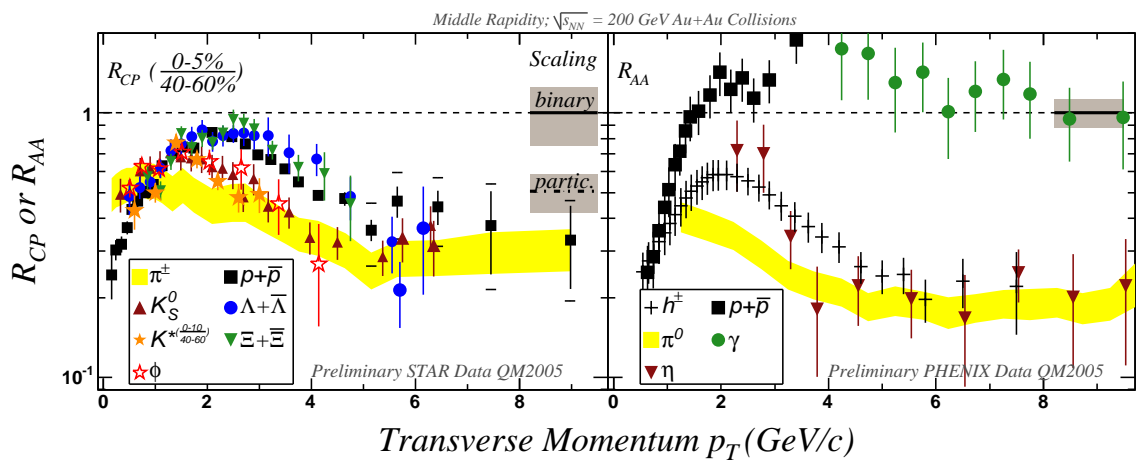}}
\vspace{-10pt}
\caption[]{ Preliminary identified particle $R_{CP}$ (left panel) and
  $R_{AA}$ (right panel). Grey bands represent the error on the
  $N_{bin}$ and $N_{part}$ calculations. For the $R_{AA}$ data, error
  bars represent both systematic and statistical uncertainties. For
  pion $R_{CP}$, the yellow band includes both systematic and
  statistical uncertainties. For proton $R_{CP}$, the systematic
  uncertainties are shown as brackets on the final five points
  (systematic uncertainties are similar for lower $p_T$ points). }
\label{rcp}
\end{figure}

Quark-number dependences at intermediate $p_T$ are also manifested in
the centrality dependence of the $p_T$ spectra. Figure~\ref{rcp} shows
the ratios $R_{CP}$~\cite{Adams:2003am,RCP,olga,dunlop} (central
Au+Au over peripheral Au+Au) in the left panel and
$R_{AA}$~\cite{phenix_spec,ph_pid_qm} (Au+Au over p+p) in the
right panel where the spectra have been scaled by the number of binary
nucleon-nucleon collisions ($N_{bin}$). For $p_T>1.5$~GeV/c, $R_{CP}$
values are grouped by hadron type (mesons vs. baryons). The larger
baryon $R_{CP}$ values indicate that baryon production increases more
quickly with centrality than meson production. This observation is
confirmed with good precision for protons and hyperons. The $\phi$ is
a particularly interesting test particle since it is a meson but is
more massive than the proton. With good precision, the $\phi$ is now
confirmed to follow the systematics of the other mesons.

Measurement of $K_S^0$ and $\Lambda+\overline{\Lambda}$ $R_{CP}$ at
$\sqrt{s_{_{NN}}}=17.2$~GeV indicate that baryon production is also
enhanced at lower center-of-mass energies~\cite{Dainese:2005vk}: with
$\Lambda+\overline{\Lambda}$ $R_{CP}$ values well above $N_{bin}$
scaling. The difference between $\Lambda+\overline{\Lambda}$ $R_{CP}$
and $K_S^0$ $R_{CP}$ is similar at all energies. This similarity is
surprising since the underlying shape of the spectrum is changing very
drastically with $\sqrt{s_{_{NN}}}$, becoming much steeper at higher
energies. It remains to be seen if any of the current theoretical
interpretations of the baryon enhancement can explain this lack of
energy dependence in the relative enhancement of baryons.

\section{Implications for non-photonic electron $R_{AA}$}

\begin{figure}[htb]
\vspace{-5pt}
\resizebox{0.5\textwidth}{!}{\includegraphics{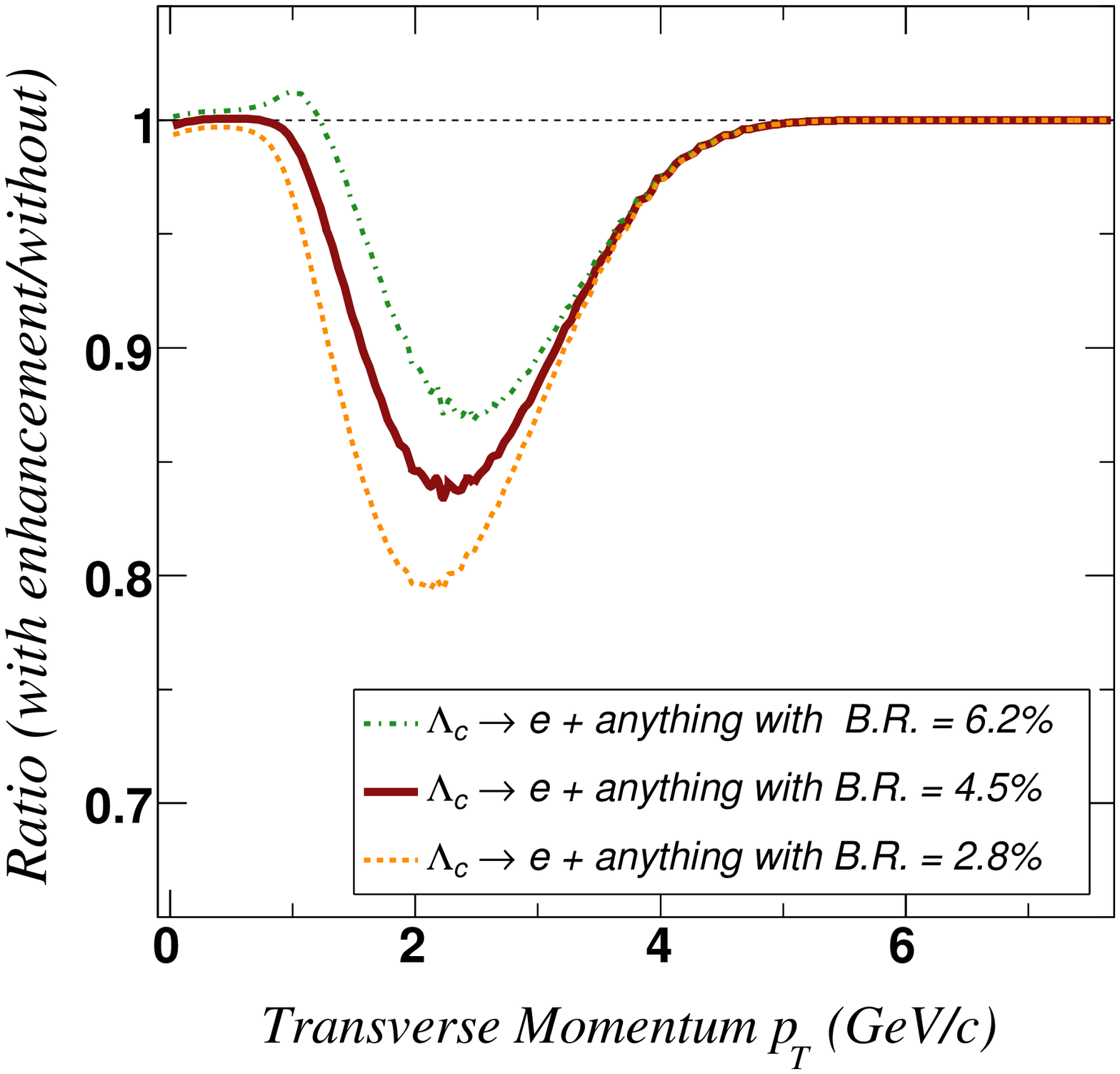}}
\resizebox{0.5\textwidth}{!}{\includegraphics{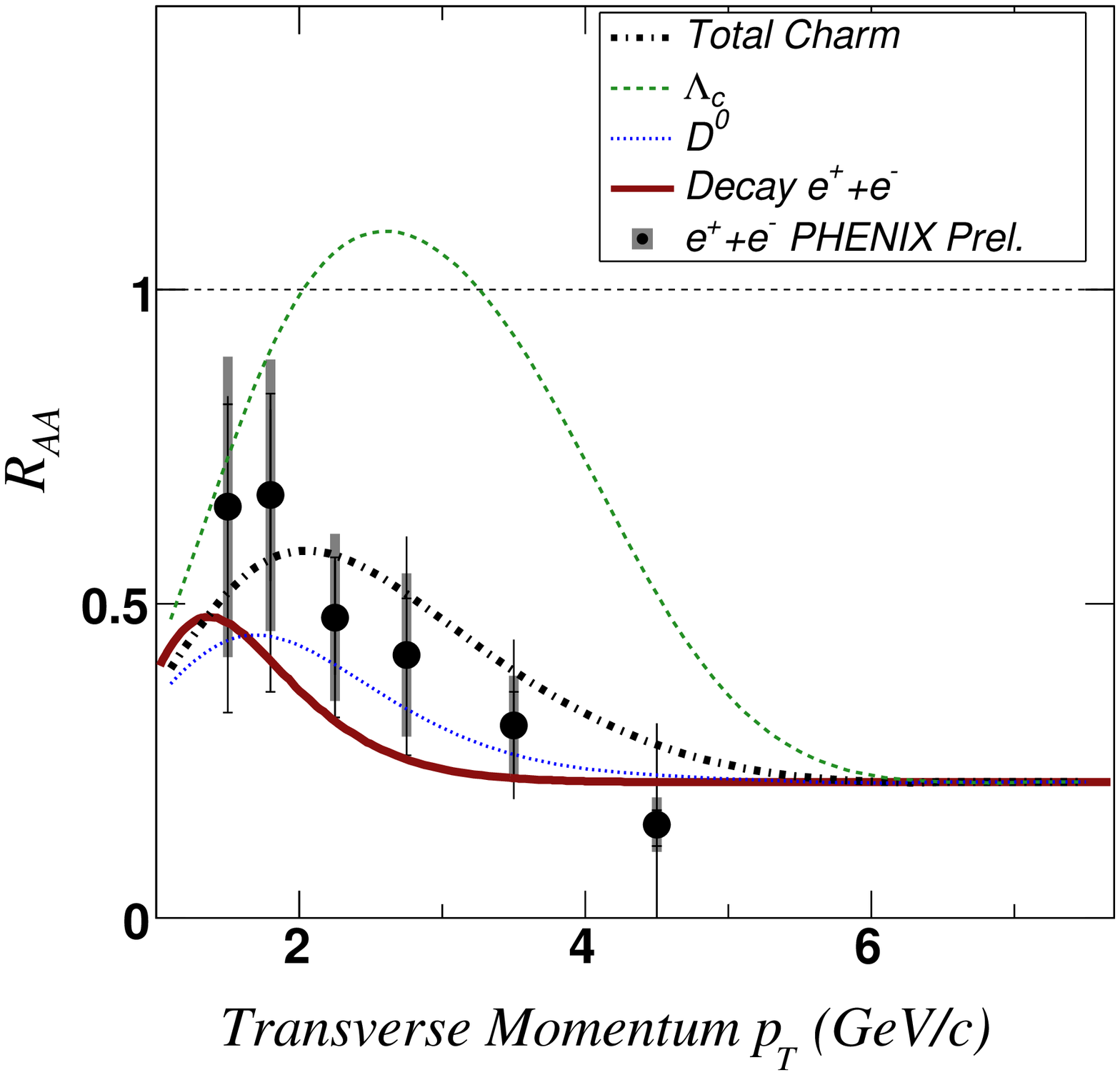}}
\caption[]{ Left panel: Electron spectrum with $\Lambda_{c}$
  enhancement divided by the spectrum without $\Lambda_{c}$
  enhancement. Right panel: $R_{AA}$ for charm hadrons and
  non-photonic electrons. The total charm spectrum in Au+Au collisions
  is scaled by the charged hadron $R_{AA}$ values. In this way the
  total charm hadron $R_{AA}$ has the same form as the charged hadron
  $R_{AA}$. The $\Lambda_{c}/D$ ratio is given the same form as the
  preliminary $\Lambda/K^0_S$ ratio. For $p_T<6$~GeV/c, the resulting
  decay electron $R_{AA}$ is smaller than either the $D$-meson or
  total charm $R_{AA}$.  }
\label{Lc}
\end{figure}

If the baryon enhancement also exists for charm hadrons, then the
non-photonic electron spectrum may be
affected~\cite{Sorensen:2005sm}. The branching ratio for $\Lambda_c
\rightarrow e$~+~\textit{anything} (4.5\% $\pm$ 1.7\%) is smaller than
that for $D^{\pm} \rightarrow e$~+~\textit{anything} (17.2\% $\pm$
1.9\%) or $D^0 \rightarrow e$~+~\textit{anything} (6.87\% $\pm$
0.28\%)~\cite{pdg}. In this case, even if charm quark production is
unchanged, increasing the $\Lambda_c/D$ ratio will lead to a reduction
in the number of observed non-photonic electrons.  In Fig.~\ref{Lc}
(left) we show the effect of a $\Lambda_{c}$ enhancement on the charm
decay electron spectrum. The ratio of two cases is taken:
$\Lambda_{c}/D$ follows the shape of the $\Lambda/K^0_S$ ratio in
Au+Au collisions, or it follows the shape of the $\Lambda/K^0_S$ ratio
in p+p collisions. A suppression of electrons from heavy flavor decays
due to the larger charm baryon-to-meson ratio in Au+Au collisions is
visible. The suppression in this figure is a result of smaller
$\Lambda_{c} \rightarrow e$~+~\textit{anything} branching ratio.  The
different curves show the experimental uncertainties on the branching
ratio~\cite{pdg}. The figure demonstrates that even if the total charm
yield follows $N_{bin}$ scaling, the non-photonic electron spectrum
can be suppressed. The magnitude of the suppression depends on the
$\Lambda_c/D$ ratio and the $\Lambda_{c} \rightarrow
e$~+~\textit{anything} branching ratio. The $\Lambda_c/D$ ratio in
Au+Au collisions is unknown but for the charm baryon-to-meson ratio
assumed here, the suppression can be as large as 20\%. In
Fig.~\ref{Lc} (right) we show the case when charm $R_{AA}$ follows
light quark $R_{AA}$ and $\Lambda_c$ is enhanced. The non-photonic
electron data are above the resulting non-photonic electron curve. If
we scale the input charm $R_{AA}$ up and calculate the $\chi^2$, we
find the smallest $\chi^2$ value for charm quark $R_{AA} = 1.35\times$
charged hadron $R_{AA}$.

\section{Interpretation of pion $R_{AA}$}

It has been proposed that the large $B/M$ values at RHIC may
be a result of the suppression of pion production coupled with novel
baryon production mechanisms scaling with $N_{bin}$ or
$N_{part}$~\cite{vitevjunctions}. The lower energy measurements call
this interpretation into question since baryon production is still
enhanced even while the suppression of mesons is much
weaker. Furthermore, it is possible that part of the pion suppression
is actually a manifestation of the baryon enhancement. Most baryons do
not decay into pions while many meson resonances do. If the relative
fraction of baryons is increasing with centrality, then that should
lead to an absence of pions from decays. Much of the difference
between Kaon and pion $R_{AA}$~\cite{ph_pid_qm,olga} may be caused by this effect: an effect
exactly analogous to that described in the previous section on
non-photonic electrons.

\section{Conclusions}

Pion, kaon, proton, and hyperon momentum-space distributions have been
measured up to $p_T \approx 10$~GeV/c. Several observations indicate
that at $p_T>6$ GeV/c, hard processes may dominate particle
production. Below this, measurements of the $B/M$ ratios, $R_{CP}$,
and $v_2$ indicate that processes beyond the reach of perturbative QCD
are prevalent. The baryon enhancement at intermediate $p_T$ is seen to
be generic to many systems and appears to be correlated with higher
multiplicity. Measurements favor pictures involving multi-parton
dynamics and point to the presence of active quark and gluon
degrees-of-freedom. A baryon enhancement in the charm sector is shown
to potentially impact the interpretation of non-photonic electron
$R_{AA}$. For the scenario presented we find that preliminary
non-photonic electron $R_{AA}$ data prefer a charm quark $R_{AA}$ 35\%
larger than light quark $R_{AA}$. We also note that the baryon
enhancement may lead to an extra suppression of pions due to the
reduction in the number of pions from resonance decays when baryon
production becomes preferred.

\section*{Acknowledgments}
I thank the conference organizers and acknowledge valuable input from
H.~Huang, Z. Xu, H.~Long, J.~Nagle, X.~Dong, N.~Xu, H-G.~Ritter,
S.~Blyth, M.~Oldenburg, J.~Chen, and Y.~Lu. I also thank the Battelle
Memorial Institute and Stony Brook University for support in the form
of the Gertrude and Maurice Goldhaber Distinguished Fellowship.

\vfill\eject
\end{document}